\documentclass[10pt]{article}
\usepackage{latexsym,graphicx}
\usepackage{epstopdf}
\usepackage{amsmath}
\usepackage{amssymb}
\newcommand{\be}{\begin{equation}}
\newcommand{\ee}{\end{equation}}
\usepackage[left=2.5cm,top=2.5cm,right=2.5cm,bottom=1.5cm]{geometry}
\begin{document}
\begin{center}
\large{\bf{Bianchi type-I Dust Filled Accelerating Brans- Dicke Cosmology}} \\
\vspace{10mm}
\normalsize{ Umesh Kumar Sharma$^1$, Gopi Kant Goswami$^2$, Anirudh Pradhan$^3$}\\
\vspace{5mm}
\normalsize{$^{1,3}$Department of Mathematics, Institute of Applied Sciences \& Humanities, G L A University,
Mathura-281 406, Uttar Pradesh, India \\
\vspace{2mm}
\normalsize{$^2$ Kalyan Post-Graduate College, Bhilai-490006, C.G., India}\\
\vspace{2mm}

$^1$E-mail: sharma.umesh@gla.ac.in  \\
\vspace{2mm}
$^2$\normalsize{Email: gk.goswami9@gmail.com}\\
\vspace{2mm}
$^3$E-mail: pradhan.anirudh@gmail.com}\\
\vspace{10mm}
\large{\bfseries{Abstract}}\
\end{center}
In this paper, spatially homogeneous and anisotropic Bianchi type-I cosmological models of Brans-Dicke theory
of gravitation are investigated. The model represents accelerating universe at present and is considered to be
dominated by dark energy. Cosmological constant $\Lambda$ is considered as a candidate for the dark energy that
has negative pressure and is responsible for the present acceleration. The derived model agrees at par with the
recent SN Ia observations. We have set BD-coupling constant $\omega$ to be ~$40000$, ~seeing the solar
system tests and evidences. We have discussed the various physical and geometrical properties of the models and
have compared  them with the corresponding relativistic models.

\section{INTRODUCTION}
 Type Ia supernovae observations \cite{ref1,ref2}, the observations of CMBR anisotropy
spectrum \cite{ref3}, large scale structure (LSS) \cite{ref4} and Planck results for CMB anisotropies \cite{ref5}
ascertain the fact that our universe is undergoing an accelerated expansion at present. It  is considered to be
dominated by dark energy that has negative pressure and is responsible for the present acceleration. As it is a well
known fact that the universe had once gone through accelerated phase during inflation for a very short period, so the
present phase may be the second attempt for it to have gone through accelerated phase. The large-scale structure surveys
and results of measurements of masses of galaxies \cite{ref6} provide the best fit value of density
parameter for matter $\Omega_{m,0} = 0.3$ and consequently $\Omega_{\Lambda,0} = 0.7$. These researches and the latest
observations explores that our universe is nearly flat.\\
The present scenario of accelerating phase of the universe and the various observational cosmological facts regarding the present
day universe are very well explained by the $\Lambda$-cold dark matter ($\Lambda$-CDM) cosmological model \cite{ref7,ref8}.
In this model, Einstein's field equations are solved for Friedmann Robertson Walker (FRW) metric in presence of positive
cosmological constant as source for dark energy along with perfect fluid distribution of the matter. It is a beauty of the
$\Lambda$-CDM model that the specific value of cosmological constant changes the decelerating phase of the universe
 into the accelerating one. The latest cosmological observations \cite{ref9,ref10} agrees with $\Lambda$-CDM  model.\\
 Spatially homogeneous and anisotropic cosmology had been a matter of interest to the cosmologist long back since 1962,
when Heckmann and Schucking \cite{ref11} wrote a chapter on anisotropic Universe. The spatially homogeneous and anisotropic
Bianchi type-I metric is often referred as Heckmann Schuking metric. It was thought that neutrino viscosity in the primordial
fire ball \cite{ref12,ref13} may create anisotropy in the Universe which dissipates
out with the advent of time. Accordingly a large number of spatially homogeneous and anisotropic solutions of Einstein's
theory have  been obtained \cite{ref14}-\cite{ref24}. Off late Wilkinson Microwave Anisotropic Probe (Bennett
et al. \cite{ref25}) also created interest in the investigation of anisotropic models of the universe. Recently, Goswami et al.
\cite{ref26}-\cite{ref30} have also developed $\Lambda$-CDM type models for Bianchi type-I anisotropic universe.\\
The fundamental and the basic philosophy behind the general theory of relativity (GTR)is that the presence of gravitational field
geometrizes the space-time. Einstein was very much impressed by the Mach philosophy that the distant background of the universe
has the impact with the local matter and that the inertial property in the matter is due to its intersection with the distant matter.
But the trouble in GTR is that it does not incorporate fully Mach's principle \cite{ref31}. A modified relativistic theory of gravitation,
 closely related to Jordan's theory \cite{ref32} and compatible to Mach's principle was developed by Brans and Dicke in 1961,
 well known as Brans-Dicke gravity \cite{ref33,ref34}. The constant coupling parameter $\omega$ and a scalar
field $\phi$ provide the intersection with the distant background of the universe. The  recent experimental
evidences \cite{ref35}$-$\cite{ref38} indicate that  the value for coupling constant $\omega$  must be higher  than  40000.
 It is found that Brans-Dicke theory goes over to GTR  when $\omega$ goes to infinity \cite{ref31}. Low energy limit of
many theories of quantum gravity (for example, superstring theory, etc.) and the cosmology have been discussed in BD-theory
by many authors \cite{ref39, ref40, ref41}. So many researchers (see \cite{ref42}-\cite{ref56} and references therein)
discussed the various burning issues like  all important features of the evolution of the universe
such as: inflation, early and late time behaviour of the universe, cosmic acceleration and structure formation, quintessence
and coincidence problem, self-interacting potential and cosmic acceleration, high energy description of dark energy in an
approximate 3-brane in Brans-Dicke theory.\\
In view of above ideas, it is worth to find out effect of cosmological constant $\Lambda$ in BD-theory of gravitation,
so that we could get history of evolution of the universe that also contain the present accelerated phase. Earlier,
Hrycyna and Lowski \cite{ref57} studied dynamical evolution of the universe in BD-theory and compared their outcome
with the corresponding results of relativistic cosmology. Dieter Lorenz-Petzold \cite{ref58} obtained Bianchi type-I BD-exact solution.
 Recently,  M. Sharif and S. Waheed \cite{ref59} and Y. Kucukakca \emph{et. al.}\cite{ref60} have developed anisotropic universe models in Brans–Dicke theory.
 Maurya et al.\cite{ref61} obtained anisotropic string cosmological models in
Brans-Dicke theory of gravitation with time dependent deceleration parameter. Off late, Goswami \cite{ref62} has developed a
$\Lambda$-CDM type cosmological model in BD-theory and has obtained a spatially flat dust filled universe in the
presence of a positive cosmological constant $\Lambda$.\\
In this paper, we have investigated spatially homogeneous and anisotropic Bianchi type-I cosmological models in Brans-Dicke
theory of gravitation. Although the format of the paper is similar to that of Goswami \cite{ref62}, yet it is a different work.
Goswami \cite{ref62} has considered a spatially flat dust filled isotropic universe where as our case is that of
a spatially flat dust filled anisotropic universe. The reason for taking anisotropy is explained above.
The outline of paper is as follows: in Section $2$, Brans-Dicke field equations for Bianchi type-I
metric are obtained. In Section $3$, we have developed a linear relationship amongst energy parameters $\Omega_m$, $\Omega_\Lambda$
and $\Omega_\sigma$. Section $4$ discusses variation of gravitational constant with red shift. In Section $5$, we have
obtained expressions for Hubble's constant, luminosity distance and apparent magnitude. We have also estimated the present
values of energy parameters and Hubble's constant. The deceleration parameter (DP), age of the universe and certain physical
properties of the universe are presented in Section $6$. Finally, conclusions are summarized in Section $7$.\\

\section{BRANS-DICKE FIELD EQUATIONS FOR BIANCHI TYPE I METRIC }
 We consider a spatially homogeneous and anisotropic  Bianchi type 1 space-time given by following metric
\begin{equation}
\label{1}
 ds^{2}= c^{2}dt^{2}- A^{2}dx^{2}-B^{2}dy^{2}-C^{2}dz^{2},
\end{equation}
where $A(t)$, $B(t)$ and $C(t)$ are scale factors along $x$, $y$ and $z$ axes.\\

The energy momentum tensor is taken as that of  perfect fluid given by following metric
\begin{equation}
\label{2}
T_{ij}=(p+\rho)u_{i}u_{j}-pg_{ij},
\end{equation}
where $g_{ij}u^{i}u^{j}=1$ and $u^{i}$ is the 4-velocity vector.\\

In co-moving co-ordinates

$$ u^{\alpha}=0,~~~~~~~~~\alpha=1,2, 3.$$

The Brans-Dicke \cite{ref33} field equations are derived from a variational principle with a Lagrangian that
generalizes the traditional one; we have included in the Brans-Dicke Lagrangian the cosmological constant for
generality (see Petrosian \cite{ref63}). Brans-Dicke field equations with cosmological constant $\Lambda$ are
obtained from following action \cite{ref64}
\begin{equation}
\label{3}
S=  \int \sqrt{-g} \left\lbrace \phi \left(
R-2\Lambda\right)+\omega\frac{\phi_{k}\phi^{k}}{\phi}+ \frac{16\pi
L_{M}}{c^{4}} \right\rbrace d^{4}x ,
\end{equation}
where $\phi$ is the scalar field representing reciprocal of varying Gravitational constant $G$, $\phi_{,i}\equiv\phi_{i}$, R is
Ricci scalar and $L_{M}$ is the matter Lagrangian.\\

 The field equations are obtained by varying the action with respect to $g_{ij}$ and $\phi$ as independent variables.
 The Brans-Dicke field equations with cosmological constant $\Lambda$ are given as follows:
\begin{equation}
\label{4}
R_{ij}-\frac{1}{2}R + \Lambda g_{ij}=- \frac{8\pi}{\phi c^4}T_{ij}
 -\frac{\omega}{\phi^2}\left( \phi_i \phi_j -\frac{1}{2}g_{ij}\phi_k \phi^k\right)
 - \frac{1}{\phi}\left( \phi_{i;j} - g_{ij} \Box \phi \right),
\end{equation}
\begin{equation}
\label{5}
(2\omega+3)\Box \phi=  \frac{8\pi  T}{ c^4}+2\Lambda \phi,
\end{equation}
where $\Box \phi = \phi^{i}_{;i}$. \\

Choosing co-moving coordinates, the field Eqs. (\ref{4}) and (\ref{5}), for the line element (\ref{1}), are obtained as
\begin{equation}
\label{6}
\frac{\ddot{B}}{B}+\frac{\ddot{C}}{C}+\frac{\dot{B}\dot{C}}{BC}=-\frac{8\pi }{\phi c^{2}} p - \frac{\omega \dot\phi^2}{2\phi^2}-
\frac{\dot{\phi}}{\phi}(\frac{\dot{B}}{B}+\frac{\dot{C}}{C}) -\frac{\ddot{\phi}}{\phi} +\Lambda c^{2},
\end{equation}
\begin{equation}
\label{7}
\frac{\ddot{A}}{A}+\frac{\ddot{C}}{C}+\frac{\dot{A}\dot{C}}{AC}= -\frac{8\pi }{\phi c^{2}} p - \frac{\omega \dot\phi^2}{2\phi^2}-
\frac{\dot{\phi}}{\phi}(\frac{\dot{A}}{A}+\frac{\dot{C}}{C}) -\frac{\ddot{\phi}}{\phi} +\Lambda c^{2},
\end{equation}
\begin{equation}
\label{8}
\frac{\ddot{A}}{A}+\frac{\ddot{B}}{B}+\frac{\dot{A}\dot{B}}{AB}= -\frac{8\pi }{\phi c^{2}} p - \frac{\omega \dot\phi^2}{2\phi^2}-
\frac{\dot{\phi}}{\phi}(\frac{\dot{A}}{A}+\frac{\dot{B}}{B}) -\frac{\ddot{\phi}}{\phi} +\Lambda c^{2},
\end{equation}
\begin{equation}
\label{9}
 \frac{\dot{A}\dot{B}}{AB}+\frac{\dot{B}\dot{C}}{BC}+\frac{\dot{C}\dot{A}}{AC}= \frac{8\pi }{\phi c^{2}}\rho-\frac{\dot{\phi}}{\phi}
 (\frac{\dot{A}}{A}+ \frac{\dot{B}}{B}+\frac{\dot{C}}{C})+ \frac{\omega \dot\phi^2}{2\phi^2} +\Lambda c^{2},
\end{equation}
\begin{equation}
\label{10}
\frac{\ddot{\phi}}{\phi}+ \frac{\dot{\phi}}{\phi}(\frac{\dot{A}}{A}+ \frac{\dot{B}}{B}+\frac{\dot{C}}{C})=
\frac{8 \pi (\rho-3p)}{(2\omega+3)c^2\phi}+\frac{2\Lambda c^2}{2\omega+3},
\end{equation}
\begin{equation}
\label{11}
  \frac{\dot{\rho}}{\rho}+\gamma\frac{\dot{(ABC)}}{ABC}=0 \;.
\end{equation}

Here  Eq. (\ref{11}) corresponds to energy conservation equation $T^{ij}_{;j}$  and  $p=(\gamma-1)\rho$  is equation of state.
 $\gamma=1$  for dust dominated universe and  $\gamma=4/3$  for radiation filled universe. Subtracting Eq. (\ref{6}) from
 Eq.(\ref{7}), Eq.(\ref{7}) from Eq.(\ref{8}) and  Eq.(\ref{6}) from Eq. (\ref{8}), we obtain
 \begin{equation}
 \label{12}
 \frac{\ddot{A}}{A}-\frac{\ddot{B}}{B}+\left(\frac{\dot{A}}{A}-\frac{\dot{B}}{B}\right)\left(\frac{\dot{C}}{C}+
 \frac{\dot{\phi}}{\phi}\right) = 0 ,
\end{equation}
\begin{equation}
\label{13}
  \frac{\ddot{B}}{B}-\frac{\ddot{C}}{C}+\left(\frac{\dot{B}}{B}-\frac{\dot{C}}{C}\right)\left(\frac{\dot{A}}{A}+
  \frac{\dot{\phi}}{\phi}\right) = 0 ,
\end{equation}
\begin{equation}
\label{14}
 \frac{\ddot{C}}{C}-\frac{\ddot{A}}{A}+\left(\frac{\dot{C}}{C}-\frac{\dot{A}}{A}\right)\left(\frac{\dot{B}}{B}+
 \frac{\dot{\phi}}{\phi}\right) = 0 .
\end{equation}

Subtracting Eq.(\ref{14}) from Eq.(\ref{12}), we get

\begin{equation}
\label{15}
\frac{\ddot{B}}{B}+\frac{\ddot{C}}{C}+\frac{2\dot{B}\dot{C}}{BC}+\left(\frac{\dot{B}}{B}+
\frac{\dot{C}}{C}\right)\frac{\dot{\phi}}{\phi} =2\frac{\ddot{A}}{A}+\left(\frac{\dot{B}}{B}+
\frac{\dot{C}}{C}\right)\frac{\dot{A}}{A}+\frac{2\dot{A}}{A}\frac{\dot{\phi}}{\phi} .
\end{equation}

This equation can be re-written in the following form

\begin{equation}
\label{16}
\frac{d}{dt}\left(\frac{\dot{(BC)}}{BC}\right)+\left(\frac{\dot{(BC)}}{BC}\right)^{2}+ \left( \frac{\dot{(BC)}}{BC}-2\frac{\dot{(A)}}{A} \right)\frac{\dot{(\phi)}}{\phi} =2\frac{d}{dt}\left(\frac{\dot{A}}{A}\right)+
2\frac{\dot{A}^{2}}{A^{2}}+\frac{\dot{A}\dot{(BC)}}{ABC}  .
\end{equation}

 Integrating this equation, we get the following first integral

 \begin{equation}
 \label{17}
 \left(\frac{\dot{(BC)}}{BC}-\frac{2\dot{A}}{A}\right)\times (ABC \phi) = L,
\end{equation}
where $L$ is constant of integration.\\

The observations show that the anisotropy existing in the past dissipates out at present. So  we take constant $L=0$.
This gives the following relationship amongst the metric coefficients
\begin{equation}
\label{18}
  A^2=BC \; .
\end{equation}
Therefore, we  may assume

\begin{equation}
\label{19}
 B =Ad\;\; \&\;\; C = \frac{A}{d},
\end{equation}
where\; d\;=\;d(t).\\
With these choices of metric coefficients, the Brans-Dicke field equations take the following form

 \begin{equation}
 \label{20}
  2{\left(\frac{\ddot{A}}{A}\right)}+\left(\frac{\dot{A}}{A} \right)^2 + \frac{\omega \dot\phi^2}{2\phi^2} +
  2\frac{\dot{\phi}}{\phi}\frac{\dot{A}}{A} +\frac{\ddot{\phi}}{\phi}  =
  -\frac{8\pi }{\phi c^{2}} p -{\left( \frac{\dot{d}}{d}\right)}^2 + \Lambda c^{2},
\end{equation}
\begin{equation}
\label{21}
\left(\frac{\dot{A}}{A} \right)^2 +\frac{\dot{\phi}}{\phi}\frac{\dot{A}}{A}-\frac{\omega \dot\phi^2}{6\phi^2}=
\frac{8\pi }{3\phi c^{2}}\rho +\frac {\Lambda c^{2}}{3}+\frac{1}{3} {\left( \frac{\dot{d}}{d}\right)}^2,
\end{equation}
\begin{equation}
\label{22}
\frac{d}{dt} \left(\frac{\dot{d}}{d}\right)+\frac{\dot{d}}{d}\left(3\frac{\dot{A}}{A}+\frac{\dot{\phi}}{\phi}\right)=0
\end{equation}
\begin{equation}
\label{23}
\frac{\ddot{\phi}}{\phi}+3\frac{\dot{A}}{A}\frac{\dot{\phi}}{\phi}=\frac{8\pi(\rho-3p)}{(2\omega+3)c^2\phi} +
\frac{2\Lambda c^2}{2\omega+3} \;.
\end{equation}
\begin{equation}
\label{24}
\frac{\dot{\rho}}{\rho} + 3\gamma\frac{\dot{A}}{A} = 0.
\end{equation}
The equation (\ref{22}) is integrable and provides following solution
\begin{equation}
\label{25}
\frac{\dot{d}}{d}= \frac{k}{A^3 \phi},
\end{equation}
where k  is constant of integration\\
Here we note that when the constant k=0,  parameter d=0 and this yields the metric coefficients
$$ A=B=C $$
 In this case our model is converted to homogeneous and  isotropic BD- model derived by  Goswami \cite{ref62}.\\
 So for anisotropic universe $ k \neq 0 $

\section{ENERGY PARAMETERS AND THEIR RELATIONSHIP FOR DUST FILLED UNIVERSE }
The universe is as at present dust dominated, so we consider $p = 0$ and $\gamma = 1$.
 Since the right hand side of  Eq. (\ref{21}) are energy densities, we define energy parameters for matter, dark energy and anisotropic
energy as follows.

\begin{equation}
\label{26}
\Omega_m=\frac{8\pi\rho}{3c^2H^2 \phi},\;\; \Omega_\Lambda=\frac{\Lambda c^2}{3H^2},
 and \;\;  \Omega_\sigma=\frac{k^2}{3H^2A^6\phi^2} \;.
 \end{equation}

We also define decelerating parameter for scale factor $a$ and scalar field $\phi$  as

 \begin{equation}
 \label{27}
  q= -\frac{\ddot{A}}{AH^2}\; and \; q_{\phi}=-\frac{\ddot{\phi}}{\phi   H^2}\;,
 \end{equation}
 where the Hubble Constant $$ H=\frac{\dot{A}}{A}$$
 Therefore, with the help of Eqs. (\ref{26}) and (\ref{27}), Eqs. (\ref{20}) to (\ref{23}) are reduced to
 \begin{equation}
 \label{28}
  -2q+1+ \frac{\omega \xi^2}{2}+2\xi-q_{\phi}=3\Omega_{\Lambda}-3\Omega_{\sigma},
  \end{equation}
 \begin{equation}
 \label{29}
 \Omega_m+\Omega_\Lambda +\Omega_\sigma =1+\xi-\frac{\omega}{6}\xi^2 ,
 \end{equation}

\begin{equation}
\label{30}
 -q_{\phi}+3\xi=\frac{3\Omega_m}{2\omega+3}+ \frac{6\Omega_{\Lambda}}{2\omega+3} ,
\end{equation}
\begin{equation}
\label{31}
 \xi= \frac{\dot{\phi}}{H\phi}.
\end{equation}

The Eqs. (\ref{28})-(\ref{31}) give rise to the following equation.

\begin{equation}
\label{32}
q - (\omega + 1)q_{\phi} + (3\omega + 2)\xi = 2 \; ,
 \end{equation}
 which has first integral as

\begin{equation}
\label{33}
(\omega+1)\frac{\dot{\phi}}{\phi}-\frac{\dot{A}}{A}=\frac{L}{\phi A^3} ,
\end{equation}
where $L$ is constant of integration. The solution Eq.(\ref{33})  has a singularity at $A=0\ and \ \phi=0$, so we
take  constant $L=0$. This gives the following power law relation between scalar field $\phi$ and scale factor$\;A$.
\begin{equation}
\label{34}
\xi= \frac{1}{\omega+1},~~\phi=\phi_0\left(\frac{A}{A_0}\right)^{\frac{1}{\omega+1}} ,
\end{equation}
where $\phi_0$ and $A_0$ are values of scalar field $\phi$ and scale factors $A$ at present.
Putting value of $\xi$ in Eq. (\ref{29}), we get following relationship amongst energy parameters.
\begin{equation}
\label{35}
\Omega_m+\Omega_\Lambda +\Omega_\sigma = 1+\frac{5\omega+6}{6(\omega+1)^2} \;.
\end{equation}
This result is analogue of the relativistic result obtained by us in \cite{ref26} in Brans-Dicke theory.
The relativistic result is as follows
\begin{equation}
\label{36}
\Omega_m+\Omega_\Lambda +\Omega_\sigma = 1 \;.
\end{equation}

\section{GRAVITATIONAL CONSTANT VERSUS REDSHIFT RELATION }

As gravitational constant $G$ is reciprocal of $\phi$ i.e.
\begin{equation}
\label{37}
G=\frac{1}{\phi}\;,
\end{equation}
and
\begin{equation}
\label{38}
\frac{A_0}{A}=(1+z),
\end{equation}
where $z$ is the red shift. \\

So, from Eqs. (\ref{34}), (\ref{37}) and (\ref{38}), we obtain
\begin{equation}
\label{39}
\frac{{G} }{G_0}= \left(1+z\right)^{\frac{1}{\omega+1}}.
\end{equation}
This result had been  obtained earlier by Goswami \cite{ref62} for spatially flat dust filled BD-universe. It is concluded that
variation of gravitational constant G over red shift z and coupling constant $\omega$ follows same pattern for both isotropic and anisotropic
BD-universe.\\
This relation ship shows that  $G$ grows toward the past and in fact it diverges at cosmological singularity.
Radar observations, Lunar mean motion and the  Viking landers on Mars \cite{ref35} suggest that rate of variation
of gravitational constant must be very much slow of order $10^{-12} year^{-1}$.  The recent experimental evidence
\cite{ref37, ref38} shows that $\omega>40000$. Accordingly, we consider large coupling constant $ \omega=40000$ in this study.

From Eqs. (\ref{33}) and (\ref{37}), the present rate of gravitational constant is calculated as
\begin{equation}
\label{40}
\left(\frac{\dot{G}}{G}\right)_0  = -\frac{1}{\omega+1} H_0 \; ,
\end{equation}
where $ H_0 \simeq  10^{-10} year^{-1}$. \\

  Eq. (\ref{39})  exhibits the fact that how $G/G_0$ varies over $\omega$.
 For higher values of $\omega$,  $G/G_0$  grows very slow over redshift, where as  for lower values of $\omega$ it grows fast.
 The variation of Gravitational constant over red shift for different $\omega$'s is already shown in fig.(1) of  ref. \cite{ref62}.

\section{EXPRESSIONS FOR HUBBLE'S CONSTANT, LUMINOSITY DISTANCE AND APPARENT MAGNITUDE }

\subsection{HUBBLE'S CONSTANT}

The energy conservation Eq. (\ref{24}) is integrable for dust filled universe, giving rise to following expression
amongst matter density $\rho$, average scale factor $a$ and the red shift $z$ of the universe
\begin{equation}
\label{41}
\rho=(\rho)_0\left(\dfrac{A_0}{A}\right)^3 = (\rho)_0\left(1+z\right)^{3}.
\end{equation}
 where we have used the relation given by the Eq. (\ref{38}). \\

 Now, using Eqs. (\ref{26}), (\ref{35}), (\ref{38}) and (\ref{41}), we get following
 expressions for Hubble's constant in terms of scale factor and redshift
\begin{equation}
\label{42}
H = \frac{H_0}{\sqrt{1+\frac{5\omega+6}{6(\omega+1)^2}}}
 \sqrt{(\Omega_m)_0 \left(\frac{A_0}{A}\right)^{\frac{3\omega+4}{\omega+1}}+
 (\Omega_{\sigma})_0 \left(\frac{A_0}{A}\right)^{2\left(\frac{3\omega+4}{\omega+1}\right)}+(\Omega_{\Lambda})_0}\, ,
\end{equation}
and
\begin{equation}
\label{43}
H = \frac{H_0}{\sqrt{1+\frac{5\omega+6}{6(\omega+1)^2}}}
 \sqrt{(\Omega_m)_0 (1+z)^{\frac{3\omega+4}{\omega+1}}+ (\Omega_{\sigma})_0 (1+z)^{2\left(\frac{3\omega+4}
 {\omega+1}\right)}+(\Omega_{\Lambda})_0}\, ,
\end{equation}
respectively.

 \subsection{ LUMINOSITY DISTANCE}

 The luminosity distance which determines flux of the source is given by
\begin{equation}
\label{44}
    D_{L}=A_{0} r (1+z),
\end{equation}
where $r$ is the spatial co-ordinate distance of a source. The luminosity distance for metric (\ref{1}) can be written as \cite{ref26}
\begin{equation}
\label{45}
D_{L}=c(1+z)\int^z_0\frac{dz}{H(z)}.
\end{equation}

Therefore, by using Eq. (\ref{43}), the  luminosity distance  $D_L$ for our model is obtained as
  \begin{equation}
  \label{46}
  D_{L}=\frac{c(1+z)\sqrt{1+\frac{5\omega+6}{6(\omega+1)^2}}}{H_{0}}\int_{0}^{z}\frac{dz}
  {\sqrt{[(\Omega_{m})_{0}(1+z)^{\frac{3\omega+4}{\omega+1}}
  + (\Omega_{\sigma})_0 (1+z)^{2\left(\frac{3\omega+4}{\omega+1}\right)}+(\Omega_{\Lambda})_{0}]}}\,.
  \end{equation}

   \subsection{ APPARENT MAGNITUDE}

  The apparent magnitude of a source of light is related to the  luminosity distance via following  expression
  \begin{equation}
  \label{47}
  m = 16.08 + 5 log_{10}\frac{H_{0}D_{L}}{.026c Mpc}.
  \end{equation}

   Using Eq. (\ref{46}), we get following expression for apparent magnitude in our model
  \begin{equation}
  \label{48}
  m=16.08+5log_{10}\left(\frac{(1+z)\sqrt{1+\frac{5\omega+6}{6(\omega+1)^2}}}{.026}\int_{0}^{z}\frac{dz}
  {\sqrt{[(\Omega_{m})_{0}(1+z)^{\frac{3\omega+4}{\omega+1}}
  + (\Omega_{\sigma})_0 (1+z)^{2\left(\frac{3\omega+4}{\omega+1}\right)}+(\Omega_{\Lambda})_{0}]}}\right).
  \end{equation}

  \subsection{ENERGY PARAMETERS AT PRESENT}

  We consider $287$ high red shift ($ 0.3 \leq z \leq 1.4$ ) SN Ia supernova data set of observed apparent
  magnitudes along with their possible error from union $2.1$ compilation  \cite{ref65}.
   In our early work  \cite {ref26} $-$ \cite{ref30}, we have presented a technique
   to estimate the present values of energy parameters $(\Omega_m)_0$, $(\Omega_\Lambda)_0$, and $(\Omega_\sigma)_0$
   by comparing the theoretical and observed results with the help of  following  $\chi^2$ formula.
   \begin{equation}
   \label{49}
  \chi_{SN}^{2}=X-\frac{Y^{2}}{Z}+log_{10}\left(\frac{Z}{2\pi}\right),
  \end{equation}
  where
  \begin{equation}
  \label{50}
  X=\overset{287}{\underset{i=1}{\sum}}\frac{\left[\left(m\right)_{ob}-\left(m\right)_{th}\right]^{2}}
  {\sigma_{i}^{2}},
  \end{equation}
  \begin{equation}
  \label{51}
  Y=\overset{287}{\underset{i=1}{\sum}}\frac{\left[\left(m\right)_{ob}-\left(m\right)_{th}\right]}
  {\sigma_{i}^{2}},
  \end{equation}
  and
  \begin{equation}
  \label{52}
  Z=\overset{287}{\underset{i=1}{\sum}}\frac{1}{\sigma_{i}^{2}}.
  \end{equation}
  Here the sums are taken over data sets of observed and theoretical values of apparent magnitude
  of $287$ supernovae.\\

  On the basis of minimum value of $\chi^2$, we get the best fit present values of $\Omega_{m}$ and
  $\Omega_{\Lambda}$. For this, the present anisotropic energy density $(\Omega_{\sigma})$ is taken
  to be very small i. e. $\Omega_{\sigma}$ = 0.0002, coupling constant $\omega$ is taken as $40000$ and
  the theoretical values are calculated from Eq. (\ref{48}). We have found that the best fit present values of $\Omega_{m}$ and
  $\Omega_{\Lambda}$  are $ (\Omega_{m})_0 = 0.294 $ and $(\Omega_{\Lambda})_0 = 0.7058$ for minimum  $\chi^2=0.6544$.\\

  The Figures $1$ and $2$  indicate how the observed values of apparent magnitudes and luminosity distances
  reach close to the  theoretical graphs for $(\Omega_\Lambda)_0$ = 0.7058, $(\Omega_m)_0 = 0.294$ and $(\Omega_\sigma)_0 = 0.0002$.

  \begin{figure}
  \begin{minipage}{0.50\textwidth}
        \centering
        \includegraphics[width=.9\textwidth]{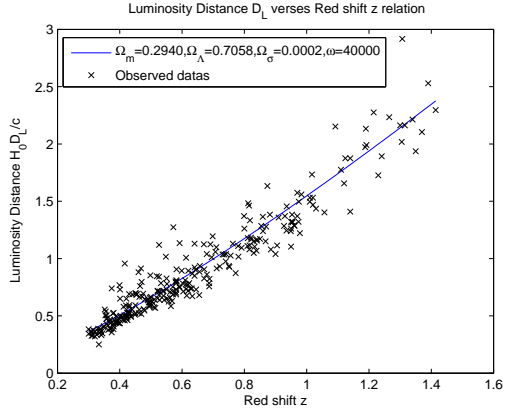}
        \caption{ Luminosity distance versus red shift best fit curve }
    \end{minipage}
     \hfill
    \begin{minipage}{0.50\textwidth}
        \centering
        \includegraphics[width=.9\textwidth]{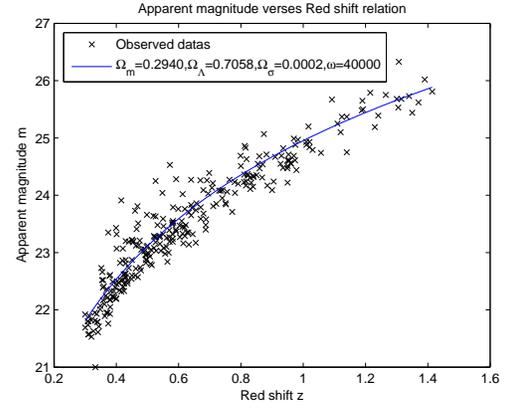}
        \caption{Apparent magnitude versus red shift best fit curve}
    \end{minipage}
    \end{figure}

 \subsection{ESTIMATION OF PRESENT VALUES OF HUBBLE'S CONSTANT $H_{0}$}
  We present a data set of the observed values of the Hubble parameters H(z) versus the red shift z with possible
  error in the form of following Table-1. These data points were obtained by various researchers from time to time,
  by using differential age approach.\\

\begin{table}
  \begin{center}
\caption[]{ Hublle's constant Table }
  	\begin{tabular}{|c|c|c|c|c|}
  		\hline $z$ & $H(z)$ &
  		$\sigma_{H}$ & Reference & Method \tabularnewline \hline \hline
  		0.07	&	69	&	19.6	&	 Moresco M. et al. \cite{ref66} 	&	DA	\tabularnewline
  		\hline									
  		0.1	&	69	&	12	&	 Zhang C. at el. \cite{ref67} 	&	DA	\tabularnewline
  		\hline									
  		0.12	&	68.6	&	26.2	&	 Moresco M. et al. \cite{ref66} 	&	DA	\tabularnewline
  		\hline									
  		0.17	&	83	&	8	&	 Zhang C. at el. \cite{ref67} 	&	DA	\tabularnewline
  		\hline
  		0.28	&	88.8	&	36.6	&	 Moresco M. et al. \cite{ref66} 	&	DA	\tabularnewline
  		\hline									
  		0.4	&	95	&	17	&	 Zhang C. at el. \cite{ref67} 	&	DA	\tabularnewline
  		\hline									
  		0.48	&	97	&	62	&	 Zhang C. at el. \cite{ref67} 	&	DA	\tabularnewline
  		\hline									
  		0.593	&	104	&	13	&	 Moresco M.  \cite{ref68}	&	DA	\tabularnewline
  		\hline									
  		0.781	&	105	&	12	&	 Moresco M.  \cite{ref68}	&	DA	\tabularnewline
  		\hline									
  		0.875	&	125	&	17	&	 Moresco M.  \cite{ref68}	&	DA	\tabularnewline
  		\hline									
  		0.88	&	90	&	40	&	 Zhang C. at el. \cite{ref67} 	&	DA	\tabularnewline
  		\hline									
  		0.9	&	117	&	23	&	 Zhang C. at el. \cite{ref67} 	&	DA	\tabularnewline
  		\hline									
  		1.037	&	154	&	20	&	 Moresco M.  \cite{ref68}	&	DA	\tabularnewline
  		\hline									
  		1.3	&	168	&	17	&	 Zhang C. at el. \cite{ref67} 	&	DA	\tabularnewline
  		\hline									
  		1.363	&	160	&	33.6	&	 Moresco M.,   \cite{ref68}&	DA	\tabularnewline
  		\hline									
  		1.43	&	177	&	18	&	 Zhang C. at el. \cite{ref67} &	DA	\tabularnewline
  		\hline									
  		1.53	&	140	&	14	&	 Zhang C. at el. \cite{ref67} 	&	DA	\tabularnewline
  		\hline									
  		1.75	&	202	&	40	&	 Zhang C. at el. \cite{ref67} 	&	DA	\tabularnewline
  		\hline									
  		1.965	&	186.5	&	50.4	&	  Stern D at el.  \cite{ref69}	&	DA	\tabularnewline
  		\hline									
  	\end{tabular}
  	\end{center}
    \end{table}

  In our model, Hubble's constant H(z) versus red shift 'z' relation Eq. (\ref{43})
  is reduced to
  \begin{equation}
  \label{53}
  H^{2}= (0.9999)H_{0}^{2}[0.294(1+z)^{3}+0.0018(1+z)^{6} +0.7058]\,
  \end{equation}
  Where we have taken $ (\Omega_{m})_0 = 0.294 $, $(\Omega_{\Lambda})_0 = 0.7058$, $\Omega_{\sigma}$ = 0.0002
  and  the coupling constant $\omega$ =  $40000$. The Hubble Space Telescope (HST) observations of
  Cepheid variables \cite{ref70} provides present value of Hubble's constant $H_0$  in the range $H_0 = 73.8 \pm 2.4 km/s/Mpc$ .
  A large number of  data sets of theoretical values of Hubble's constant H(z) versus z, corresponding to $H_0$
  in the range ($ 69 \leq H_0 \leq 74$ ) are obtained by using equation (\ref{53}). It should be noted
   that the red shift $z$  are taken from Table-1 and each data set will consist of $19$ data points.\\

  In order to get the best fit theoretical data set of Hubble's constant $H(z)$ versus $z$, we calculate
  $\chi^{2}$  by using following statistical formula.

  \begin{equation}
  \label{54}
  \chi_{SN}^{2}= \frac{X}{19},
  \end{equation}
  where
  \begin{equation}
  \label{55}
   X = \overset{19}{\underset{i=1}{\sum}}\frac{\left[\left(H\right)_{ob}-\left(H\right)_{th}\right]^{2}}
  {\sigma_{i}^{2}}.
  \end{equation}
  Here the sums are taken over data sets of observed and theoretical values of Hubble's constants. The observed
  values are  taken from Table-1 and  theoretical values are calculated from Eq. (\ref{49}). \\

  Using Eqs. (\ref{54})-(\ref{55}), we have found that best fit value of Hubble's constant $H_0$ is $71.9$ for minimum
  $\chi^2 = 0.5865$
   Figure $3$ shows the dependence of Hubble's constant with
  red shift. Hubble's observed data points are closed to the graph corresponding
  to $(\Omega_\Lambda)_0$ = $0.7058$, $(\Omega_m)_0 = 0.294$ and $(\Omega_\sigma)_0 = 0.0002$. This validates the
  proximity of observed and theoretical values.

     \begin{figure}
     \centering
     \includegraphics[width=.60\textwidth]{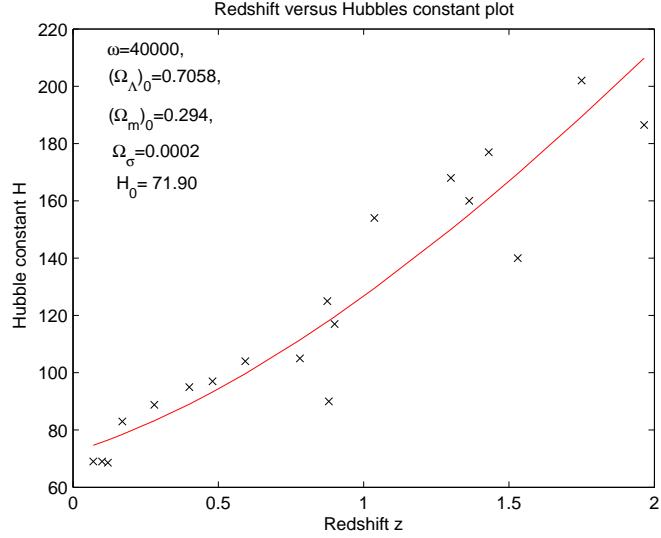}
     \caption{Variation of Hubble's constant over red shift;  Best fit curve}
     \end{figure}

 \section{CERTAIN PHYSICAL PROPERTIES OF THE UNIVERSE }

  \subsection{MATTER, DARK AND ANISOTROPIC ENERGY DENSITIES}

  The matter and  dark energy densities of the universe are related to the energy parameters through following equation
  \begin{equation}
  \label{56}
  \Omega_{m}=\frac{(\rho)_m}{\rho_{c}}, ~ ~
  \Omega_{\Lambda}=\frac{\rho_{\Lambda}}{\rho_{c}}, ~~
  \Omega_{\sigma}=\frac{(\rho)_\sigma}{\rho_{c}},
  \end{equation}
  where
  \begin{equation}
  \label{57}
  \rho_{c}=\frac{3c^{2}H^{2}}{8\pi G}=\frac{3c^{2}\phi H^{2}}{8\pi }.
  \end{equation}
 So,
  \begin{equation}
  \label{58}
   (\rho_m)_0=(\rho_{c})_0 (\Omega_{m})_0= ,\; (\rho_{\Lambda})_0=(\rho_{c})_0(\Omega_{\Lambda})_0.
  \end{equation}
  Now the present valu of $\rho_{c}$ is obtained as

  $$  (\rho_{c})_0=\frac{3c^{2}H_{0}^{2}}{8\pi G} = 1.88\; h^2_0 \times 10^{-29}\; gm/cm^3 .$$

  The estimated value of $h_0= 0.719$. Therefore, the present value of matter  and dark energy densities are given by
  \begin{equation}
  \label{59}
  (\rho_{m})_{0}=   0.5527 h^2_0\times10^{-29}gm/cm^{3},
  \end{equation}
  \begin{equation}
  \label{60}
   (\rho_{\Lambda})_{0}=\rho_c(\Omega_{\Lambda})_{0} = 1.3269 h^2_0\times10^{-29}gm/cm^{3},
   \end{equation}
    and
   \begin{equation}
   \label{61}
  (\rho_{\sigma})_{0}= 0.0003 h^2_0\times10^{-29}gm/cm^{3}.
  \end{equation}

   Here, we have taken $$ (\Omega_{m})_0= 0.294\;  (\Omega_{\Lambda})_0= 0.7058  \; \& \; (\Omega_{\sigma})_0= 0.0002 . $$

   General expressions for matter and dark energies are given by

   \begin{equation}
   \label{62}
   \rho=(\rho)_0\left(\dfrac{a_0}{a}\right)^3=(\rho)_0\left(1+z\right)^3,
   \end{equation}
   \begin{equation}
   \label{63}
  (\rho_{\Lambda})=\rho_c\;\Omega_{\Lambda},
  \end{equation}
  and
  \begin{equation}
  \label{64}
  (\rho_{\sigma})=(\rho_{\sigma})_0\left(\dfrac{a_0}{a}\right)^6=(\rho)_0\left(1+z\right)^6.
  \end{equation}
  From above, we observe that the current matter and dark energy densities are very close to the values predicted by
  the various surveys described in the introduction.

    \subsection{AGE OF THE UNIVERSE}

    By using the standard formula
    \begin{equation*}
    t = \intop_{0}^{t}dt=\intop_{0}^{A}\frac{dA}{AH},
    \end{equation*}
    we obtain the values of $t$ in terms of scale factor and redshift respectively
    \begin{equation}
    \label{65}
    t  =  \intop_{0}^{A}\frac{\sqrt{1+\frac{5\omega+6}{6(\omega+1)^2}}dA}{AH_0\sqrt{(\Omega_m)_0 (\frac{A_0}{A})^{\frac{3\omega+4}
    {\omega+1}} +(\Omega_{ \sigma})_0(\frac{A_0}{A})^{\frac{2(3\omega+4)}{\omega+1}}+(\Omega_{\Lambda})_0}}
    \end{equation}
    \begin{equation}
    \label{66}
    t  =  \intop_{0}^{z}\frac{\sqrt{1+\frac{5\omega+6}{6(\omega+1)^2}}dz}{(1+z)H_0\sqrt{(\Omega_m)_0 (1+z)^{\frac{3\omega+4}{\omega+1}}+
    (\Omega_{ \sigma})_0(1+z)^{\frac{2(3\omega+4)}{\omega+1}}+(\Omega_{\Lambda})_0}}
    \end{equation}

   For $\omega=40000$, $(\Omega_\Lambda)_0$ = $0.7058$, $(\Omega_m)_0 = 0.294$ and $(\Omega_\sigma)_0 = 0.0002$, Eq.(\ref{66})
   gives  $t_0\rightarrow 0.9677 H_0^{-1}$ for high redshift. This means that  the present age of
   the universe is $ t_{0}$ = $ 13.1392$  Gyrs as per our model.  From WMAP data, the empirical value of present age of
   universe is $13.73_{-.17}^{+.13}Gyrs$ which is closed to
    present age of universe, estimated by us in this paper. \\

     Figures $4$ shows the variation of time over red shift. At $z=0.8902$ curve becomes stationary. This provides present
     age of the universe  This also indicated the consistency with recent observations.

     \begin{figure}
     \centering
     \includegraphics[width=.60\textwidth]{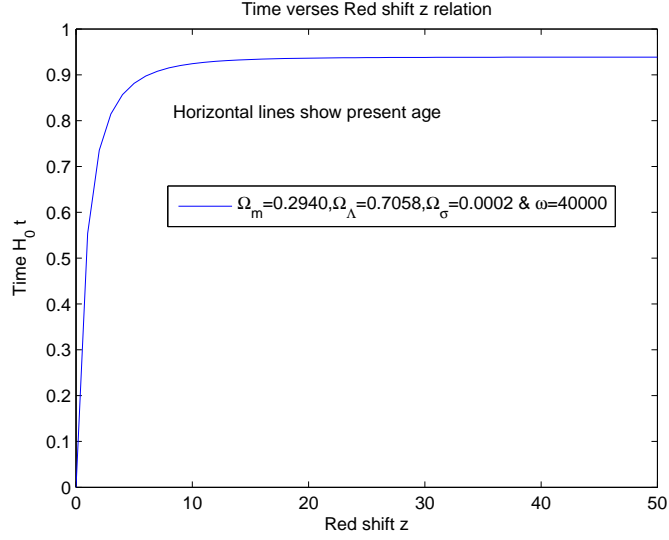}
     \caption{Variation of red shift over time and age of the universe}
     \end{figure}

   \subsection{DECELERATION PARAMETER}

  From Eqs. (\ref{28}), (\ref{32}) and (\ref{34}), we obtain the expressions for DP as
   \begin{equation}
   \label{67}
   q =\frac{\omega+2}{2(\omega+1)}- \frac{3(\omega+1)}{2\omega+3}\;(\Omega_{\Lambda} - \Omega_{\sigma}).
   \end{equation}

  Using Eqs. (\ref{26}), (\ref{34}), (\ref{42}) and (\ref{43}) in Eq. (\ref{67}), we get following expression for
  deceleration parameter
\begin{equation}
\label{68}
q =\frac{\omega+2}{2(\omega+1)} - \frac{3(\omega+1)}{2\omega+3}\;\frac{\left(1+\frac{5\omega+6}{6(\omega+1)^2}\right)
\left[\left(\Omega_\Lambda \right)_0-\left(\Omega_\sigma \right)_0(\frac{A_0}{A})^{6 + \frac{2}{(\omega + 1)}} \right]}
{\left((\Omega_m)_0\left(\frac{A_0}{A}\right)^{\frac{3\omega+4}{\omega+1}}+
(\Omega_{\sigma})_0\left(\frac{A_0}{A}\right)^{\frac{2(3\omega+4)}{\omega+1}}+(\Omega_{\Lambda})_0\right)}.
\end{equation}
In term of redshift, $q$ is given by
\begin{equation}
\label{69}
q =\frac{\omega+2}{2(\omega+1)} - \frac{3(\omega+1)}{2\omega+3}\;\frac{\left(1+\frac{5\omega+6}{6(\omega+1)^2}\right)
\left(\left(\Omega_\Lambda \right)_0-\left(\Omega_\sigma \right)_0(1+z)^{6 + \frac{2}{(\omega + 1)}} \right)}
{\left((\Omega_m)_0(1+z)^{\frac{3\omega+4}{\omega+1}}+ (\Omega_{\sigma})_0(1+z)^{\frac{2(3\omega+4)}{\omega+1}}+(\Omega_{\Lambda})_0\right)}.
\end{equation}
For $\omega=40000$, the decelerating parameter is obtained as
\begin{equation}
\label{70}
q = 0.5 - 1.5\;\left[(\Omega_\Lambda)_0 - (\Omega_\sigma)_0 (1+z)^6 \right] \frac{1}
{\left((\Omega_m)_0\left(1+z\right)^{3}+ (\Omega_{\sigma})_0\left(1+z\right)^{6}+(\Omega_{\Lambda})_0\right)}
\end{equation}

As the present phase ($z=0$) of the universe is accelerating $q \leq 0\; i.e. \; \frac{\ddot{a}}{a}\geq 0 $ , so
we must have
\begin{equation}
\label{71}
(\Omega_\Lambda)_0 \geq\dfrac{(2\omega+3)(\omega+2)}{6(\omega+1)^2}+ (\Omega_\sigma)_0.
\end{equation}

For $\omega=40000$ and $(\Omega_\sigma)_0= 0.0002$ the minimum value of $(\Omega_\Lambda)_0$ is given by
$(\Omega_\Lambda)_0 \geq  0.3335$ which is consistent with the present observed value of $(\Omega_{\Lambda})_{0}=0.7058$. \\

 Putting $z=0$ in Eq. (\ref{70}), the present value of deceleration constant is obtained as

 \begin{equation}
\label{72}
  q_0 = -0.5584\; .
 \end{equation}

 The Eq. (\ref{70}) also provides

\begin{equation}
\label{73}
z_{c} \cong 0.6805 \; at \; q=0
\end{equation}

Therefore, the universe attains to the accelerating phase when $z<z_c$. \\

Converting redshift into time from Eq. (\ref{73}), the value of $z_c$ is reduced to
\begin{equation}
\label{74}
z_{c}=0.6805 \thicksim    0.4442 H^{-1}_0 yrs\thicksim   6.0337\times10^9 yrs .
\end{equation}
 So, the acceleration must have begun in the past at $  6.0337\times10^9 yrs$
 before from present. The figure $6$ shows how deceleration parameter increases
 from negative to positive over red shift which means that in the past universe was decelerating and at a instant
 $z_{c} \cong 0.6749$, it became stationary there after it goes on accelerating.

 \begin{figure}
    \centering
    \includegraphics[width=0.60\textwidth]{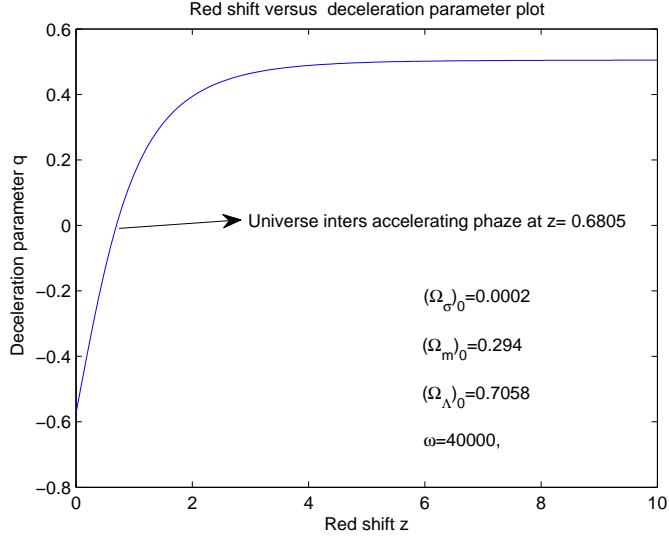}
    \caption{Variation of decelerating parameter over red shift. It represents accelerating universe at present.}
\end{figure}

\subsection{SHEAR SCALAR}

The shear scalar is given by
\begin{equation}
\label{75}
\sigma^2=\frac{1}{2}\sigma_{ij}\sigma^{ij} ,
\end{equation}
where
\begin{equation}
\label{76}
 \sigma_{ij}= u_{i;j}-\Theta(g_{ij}-u_iu_j)
\end{equation}

In our model
\begin{equation}
\label{77}
    \sigma^2= \frac{\dot{d}^2}{d^2}=
    \frac{k^2}{\phi A^6}=(\Omega_\sigma)_0H^2_0(1+z)^\frac{2(3\omega +4)}{\omega+1}
\end{equation}
From Eq. (\ref{77}), it is clear that shear scalar vanishes as $A\rightarrow\infty$.

\subsection{RELATIVE ANISOTROPY}
The relative anisotropy is given by
\begin{equation}
\label{78}
  \frac{\sigma^2}{\rho_m}=\frac{3(\Omega_\sigma)_0H^2_0(1+z)^{\frac{3\omega+5}{\omega+1}}}{(\rho_c)_0(\Omega_m)_0}
\end{equation}
This follows the same pattern as shear scalar. This means that relative anisotropy
decreases over scale factor i.e. time.

 \section{CONCLUSION}
\begin{table}
 \begin{center}
\caption[]{Cosmological  parameters at present}
\;
    \begin{tabular}{|c|c|}
        \hline
        Cosmological Parameters & Values at Present \\
        \hline
         BD coupling constant $\omega$ & 40000\\
         Dark energy parameter $(\Omega_\Lambda)_0$ & 0.7058 \\
         Dust energy parameter $(\Omega_m)_0$ & 0.294\\
         Dust energy parameter $(\Omega_\sigma)_0$ & 0.0002\\
         Hubble's constant $H_0$ & 71.90\\
         Deceleration parameter  $q_0$ & $ -0.5584$.\\
         Dust energy density $(\rho_{m})_{0}$ & $ 0.0.5527 h_0^2\times10^{-29}gm/cm^{3}$  \\
         Dark energy density $(\rho_{\Lambda})_{0}$ & $ 1.3269 h_0^2\times10^{-29}gm/cm^{3}$\\
         Anisotropic energy density $\rho_{\sigma})_{0}$ & $0.0003 h_0^2\times10^{-29}gm/cm^{3}$\\
         Age of the universe $t_{0}$ & $ 13.1392~~ Gyr$\\
        \hline
    \end{tabular}
 \end{center}
\end{table}
 We summarize our results by presenting Table-2 which displays the values of cosmological parameters at
 present obtained by us. We have found  that the acceleration would have begun in the past at
 $  6.0040\times10^9 yrs$ before from present. These results are in good agreements with the various surveys
 described in the introduction.

 \section{DISCLOSURE STATEMENT}
 The authors are not aware of any affiliation, membership, funding, or financial holding that might be perceived as affecting
 the objectivity of this paper.

 \section*{ACKNOWLEDGEMENT}
This work is  supported by the CGCOST  Research Project 789/CGCOST/MRP/14. The authors are  thankful to IUCAA, Pune, India
for providing facility and support where part of this work was carried out during a visit. Authors are also thankful to Prof J. V.
Narlikar, IUCAA for looking at the paper and making useful comment in first draft. The authors are grateful to the anonymous
referee for valuable comments to improve the quality of manuscript


\begin{thebibliography}{99}

\bibitem {ref1}
S. Perlmutter et al. (Supernova Cosmology Project collaboration), Astrophys. J. {\bf 517}, 565 (1999); [astro-ph/9812133].
\bibitem {ref2}
A. G. Riess et al. (Spurnova Serach Team collaboration), Astron. J. {\bf 116}, 1009 (1998); [arXiv:astro-ph/9805201].
\bibitem {ref3}
D. N. Spergel et al. (WMAP collaboration), Astrophys. J. Suppl. {\bf 148}, 175 (2003); [astro-ph/0302209].
\bibitem {ref4}
M. Tegmark et al. (SDSS collaboration), Phys. Rev. D {\bf 69}, 103501 (2004); [astro-ph/0310723].
\bibitem {ref5}
P. A. R. Ade et al. (Planck Collaboration), A \& A {\bf 594}, A13 (2016); arXiv:1502.01589[astro-ph.Co].
\bibitem {ref6}
R. K. Knop et al., Ap. J. {\bf 598}, 102 (2003); [arXiv:astro-ph/0309368].
\bibitem{ref7}
E. J. Copeland, M.Sami, and S. Tsujikawa, Int. J. Mod. Phys. D {\bf 15}, 1753 (2006).
\bibitem{ref8}
Ø. Grøn and  S. Hervik, {\it Einstein's general theory of relativity with modern applications in cosmology} (Springer 2007).
\bibitem {ref9}
K. Abazajian et al. (SDSS Collaboration), Astron. J. {\bf 128}, 502 (2004).
\bibitem {ref10}
V. Sahni and A. A.Starobinsky, Int. J. Mod. Phys. D {\bf 9}, 373 (2000).
\bibitem {ref11}
O. Heckmann and E. Schucking, {\it Relativistic Cosmology In Gravitation: An Introduction to current research},
ed L. Witten, Chap XI ( Willey, New York 1962) p. 438.
\bibitem {ref12}
A. G. Doroshkevich and Ya. B. Zeldovich, Sov. Phys. JETP Lett. {\bf 5}, 3 (1967).
\bibitem {ref13}
C. W. Misner, Phys. Rev. Lett. {\bf 19}, 533 (1967).
\bibitem {ref14}
G. F. R. Ellis and M. A. H. MacCallum, Communi. Math. Phys. {\bf 12}, 108 (1969).
\bibitem {ref15}
R. A. Matzner, Astrophys. J. {\bf 157}, 1085 (1969).
\bibitem {ref16}
V. B. Johri and G. K. Goswami, Aust. J. Phys. {\bf 34}, 261 (1981).
\bibitem {ref17}
V. B. Johri and G. K. Goswami, Aust. J. Phys. {\bf 34}, 235  (1983).
\bibitem {ref18}
A. Pradhan and H. Amirhashchi, Mod. Phys. Lett. A {\bf 26}, 2261 (2011); [arXiv:1110.1019[physics.gen-ph]].
\bibitem {ref19}
A. Pradhan, Res. Astron. Astrophys. {\bf 13}, 139 (2013); arXiv:1209.4826[physics.gen-ph].
\bibitem {ref20}
A. Pradhan, A. K. Pandey, and R. K. Mishra, {\bf 88}, 757 (2014).
\bibitem {ref21}
A. Pradhan and B. Saha, Phys. Parti. Nuclei, {\bf 46}, 310 (2015).
\bibitem {ref22}
D. C. Maurya, R. Zia, and A. Pradhan, Int. J. Geom. Meth. Mod. Phys. {\bf 14}, 1750077 (2017).
\bibitem {ref23}
G. P. Singh, B. K. Bishi and P. K. Sahoo, Int. J. Geom. Meth. Mod. Phys. {\bf 13}, 1650058 (2016).
\bibitem {ref24}
P. K. Sahoo, P. Sahoo, and B. K. Bishi, Int. J. Geom. Meth. Mod. Phys. {\bf 14}, 1750097 (2017).
\bibitem {ref25}
C. L. Bennett et al., The Astrophys. J. Supplement {\bf 148}, 1 (2003); [astro-ph/0302207].
\bibitem{ref26}
G. K. Goswami, M. Mishra, and A. K. Yadav, Int. J. Theor. Phys. {\bf 54}, 315 (2015).
\bibitem{ref27}
G. K. Goswami, A. K. Yadav, R. N. Dewangan, and A. Pradhan, Astrophys. Space Sci. {\bf 361}, 47 (2016).
\bibitem{ref28}
G. K. Goswami, R. N. Dewangan, and A. K. Yadav, Astrophys. Space Sci. {\bf 361}, 119 (2016).
\bibitem{ref29}
G. K. Goswami, A. K. Yadav and R. N. Dewangan, Int. J. Theor. Phys. {\bf 55}, 4651 (2016).
\bibitem{ref30}
G. K. Goswami, R. N. Dewangan and A. K. Yadav, Gravitation \& Cosmology {\bf 22}, 388 (2016).
\bibitem {ref31}
S. Weinberg, {\it Gravitation and Cosmology: Principle and Application of the General Theory of Relativity} (Wiley, NY 1972).
\bibitem {ref32}
P. Jordan, Schwerkraft and Weltall (Friedrick Vieweg and Sohn, Braunschweig, 1955).
\bibitem {ref33}
C. H. Brans and R. H. Dicke, Phys. Rev. A, Ser-2 {\bf 124}, 925 (1961).
\bibitem {ref34}
R. H. Dicke, Phys. Rev. A, Ser-2 {\bf 125}, 2163 (1962).
\bibitem{ref35}
J. V. Narlikar, {\it An Introduction to Cosmology} (Cambridge University Press 2002), p 483.
\bibitem {ref36}
R. D. Reasenberg et al., Astrophys. J. {\bf 234}, L219 (1979).
\bibitem{ref37}
B. Bertotti et al., Nature  {\bf 425}, 374 (2003).
\bibitem{ref38}
A. D. Felice et al., Phys. Rev. D  {\bf 74}, 103005 (2006).
\bibitem {ref39}
V. Faraoni, Phy. Rev. D {\bf 70}, 047301 (2004).
\bibitem {ref40}
E. Elizalde, S. Nojiri, S. D. Odintsov, and Peng Wang, Phy. Rev. D {\bf 70}, 103504 (2005).
\bibitem {ref41}
S. Nojiri and S. D. Odintsov, Gen. Relativ. Gravit. {\bf 38}, 1285 (2006).
\bibitem {ref42}
A. Errahmani and T. Ouali, preprint (2008), arXiv:0706.0115[gr-qc].
\bibitem {ref43}
W.-Q. Yang et al., Mod. Phys. Lett. {\bf 26}, 191 (2011).
\bibitem {ref44}
M. Sahraee and M. R. Setare, Int. J. Mod. Phys. D {\bf 25}, 1650097 (2016).
\bibitem {ref45}
B. K. Sahoo and L. P. Singh, Mod. Phys. Lett. {\bf 18}, 2725 (2003).
\bibitem {ref46}
M. Arik and M. C. Calik, Mod. Phys. Lett. {\bf 21}, 1241 (2006).
\bibitem {ref47}
El-Nabulsi A. Rami, Mod. Phys. Lett. {\bf 23}, 401 (2008).
\bibitem {ref48}
M. K. Mak and T. Harko, Int. J. Mod. Phys. D {\bf 12}, 925 (2003).
\bibitem {ref49}
E. Elizalde, S. Nojiri, and S. D. Odintsov, Phy. Rev. D {\bf 70}, 043539 (2004) 043539.
\bibitem {ref50}
S. Capozziello, R. De Ritis, C. Rubano, and P. Scudellaro, Int. J. Mod. Phys. D {\bf 05}, 85 (1996).
\bibitem {ref51}
S. Chakraborty, N. C. Chakraborty, and Ujjal Debnath, Mod. Phys. Lett. {\bf 18}, 1549 (2003).
\bibitem {ref52}
F. Rahaman and P. Ghosh, Mod. Phys. Lett. A {\bf 23}, 2763 (2008).
\bibitem {ref53}
S. Capozziello and G. Lambiase, Mod. Phys. Lett. {\bf 14}, 2193 (1999).
\bibitem {ref54}
A. Beesham, Mod. Phys. Lett. {\bf 13}, 805 (1998).
\bibitem {ref55}
S. Capozziello and G. Lambiase,  Mod. Phys. Lett. {\bf 30}, 1540032 (2015).
\bibitem {ref56}
A. Chand, R. K. Mishra, and A. Pradhan, Astrophys. Space Sci. {\bf 361}, 81 (2016).
 \bibitem{ref57}
O. Hrycyna and M. S. Lowski, Phys. Rev. D. {\bf 88}, 064018 (2013).
\bibitem{ref58}
Dieter Lorenz-Petzold, Phys. Rev. D. {\bf 29}, 2399 (1984).
\bibitem{ref59}
M. Sharif and S. Waheed, European Phys. Jour. C {\bf 72}, 1876 (2012).
\bibitem{ref60}
 Y. Kucukakca et al., Gen. Relativ. Gravit. {\bf 44}, 1893 (2012).
\bibitem{ref61}
D. C. Maurya, R. Zia, and A. Pradhan, J. Experim. Theor. Phys. {\bf 123}, 617 (2016).
\bibitem{ref62}
G. K.Goswami, Res. Astron. Astrophys. {\bf 17}, 1 (2017).
\bibitem{ref63}
V. Petrosian, in M. S. Longair (ed.), Confrontation of cosmological theories with observational data, {\it IAU Symp.} {\bf 63},
(D. Reidel Publ. Co., Dordrecht, 1974).
\bibitem{ref64}
F. Occhionero and F. Vagnetti, Astron \& Astrophys. {\bf 44}, 329 (1975).
\bibitem{ref65}
N. Suzuki et al., Astrophys. J. {\bf 746}, 85 (2012).
\bibitem{ref66}
M. Moresco et al., JCAP {\bf 1208}, 006 (2012); [arXiv:1201.3609].
\bibitem{ref67}
C. Zhang et al., Res. Astron. Astrophys. {\bf 14}, 1221 (2014), [arXiv:1207.4541[astro-ph.CO]].
\bibitem{ref68}
M. Moresco, Mon. Not. R. Astron. Soc. {\bf 450}, L16 (2015).
\bibitem{ref69}
D. Stern et al., JCAP {\bf 2010}, 008 (2010).
\bibitem{ref70}
V. Sahni, A. Shafieloo, and A. A. Starobinsky, Astrophys. J. Lett. { \bf 793 }, L40 (2014).
\end{thebibliography}
\end{document}